\documentclass[aps,prb,reprint,twocolumn,superscriptaddress,showpacs]{revtex4-1}
\usepackage{graphicx}
\usepackage{mathrsfs}
\usepackage{bm}
\usepackage{amsmath}
\usepackage{dcolumn}
\usepackage{epstopdf}
\usepackage{dsfont}
\usepackage{amssymb}
\usepackage{tabularx}
\usepackage{array}
\usepackage{float}
\usepackage{color}
\usepackage{epstopdf}
\usepackage{mathrsfs}
\usepackage[colorlinks, linkcolor=blue,anchorcolor=blue,citecolor=blue,urlcolor=blue]{hyperref}

\begin{document}
\title{Coexistence of four-band nodal rings and triply-degenerate nodal points in centrosymmetric metal diborides }

\author{Xiaoming Zhang}
\affiliation{Research Laboratory for Quantum Materials, Singapore University of Technology and Design, Singapore 487372, Singapore}

\author{Zhi-Ming Yu}
\affiliation{Research Laboratory for Quantum Materials, Singapore University of Technology and Design, Singapore 487372, Singapore}

\author{Xian-Lei Sheng}
\affiliation{Research Laboratory for Quantum Materials, Singapore University of Technology and Design, Singapore 487372, Singapore}
\affiliation{Department of Applied Physics, Beihang University, Beijing 100191, China}

\author{Hui Ying Yang}
\email{yanghuiying@sutd.edu.sg}
\affiliation{Research Laboratory for Quantum Materials, Singapore University of Technology and Design, Singapore 487372, Singapore}

\author{Shengyuan A. Yang}
\email{shengyuan\_yang@sutd.edu.sg}
\affiliation{Research Laboratory for Quantum Materials, Singapore University of Technology and Design, Singapore 487372, Singapore}


\begin{abstract}
Topological metals with protected band-crossing points have been attracting great interest. Here we report novel topological band features in a family of metal diboride materials. Using first-principles calculations, we show that these materials are metallic, and close to Fermi level, there appears coexistence of one pair of nodal rings and one pair of triply-degenerate nodal points (TNPs). The nodal ring here is distinct from the previously studied ones in that its formation requires four entangled bands, not just two as in previous cases, hence it is termed as a four-band nodal ring (FNR). Remarkably, we show that FNR features Dirac-cone-like surface states, in contrast to the usual drumhead surface states for two-band nodal rings. Due to the presence of inversion symmetry, the TNP here is also different from those discussed previously in inversion-asymmetric systems. Especially, when spin-orbit coupling is included, the TNP here transforms into a novel Dirac point that is close to the borderline between the type-I and type-II Dirac point categories. We discuss their respective symmetry protections, and construct effective models for their characterization. The large linear energy range ($> 2$ eV) in these materials should facilitate the experimental detection of the signatures of these nontrivial band crossings.
\end{abstract}


\maketitle
\section{Introduction}

Topological states of matter have been attracting significant interest in current condensed matter physics research. There have been extensive studies of topological insulating materials~\cite{Hasan2010,Qi2011}, in which the bulk is insulating and carries certain symmetry-protected topological order stablized by the bandgap, dictating the existence of robust gapless excitations on the boundary of the system. Soon, it was realized that similar topological characterization could also be applied to metallic states~\cite{Volovik,Zhao2013,Chiu2016}. Such topological metals or semimetals possess nontrivial band-crossings in the bulk close to the Fermi energy, around which the quasiparticle excitations could behave drastically different from the usually Schr\"{o}dinger-type fermions. For example, in a Weyl semimetal~\cite{Wan,Murakami,Burkov}, the bands cross at isolated Weyl points, around which the low-energy quasiparticles are described by Weyl fermions with a definite chirality~\cite{NN}. With certain crystalline symmetry, two Weyl points with opposite chirality can be stabilized at the same $k$-point, forming a Dirac point that features massless Dirac fermions~\cite{Young,ZWang1,ZWang2,BJYang}. Several Weyl and Dirac materials have been proposed and a few have been identified in experiment~\cite{ZKLiu2014,Borisenko,Weng2015,Huang2015,Lv2015,Xu2015}.

Currently, the exploration is extended to novel types of band-crossings beyond the Weyl and Dirac paradigm. The novelty could be either from the dimensionality of the band-crossing, or from its degeneracy. Regarding the former, materials with one-dimensional (1D) band-crossings---nodal rings---have emerged as an interesting topic~\cite{Fang2016,Yang2014,WengRing,YPChen,Yu2015,Kim2015,Cava2015,Fang2015,Chan2016,Li2016,Bian2016,Schoop}. In the identified nodal ring materials, the ring is formed by the crossing between two bands, with linear dispersions along transverse directions. On the system boundary, it features the drumhead surface states inside the surface-projected nodal ring. Whereas for the latter, crossing-points with three-, six-, or eight-fold degeneracies have been proposed, mostly resulting from nonsymmorphic space-group symmetries~\cite{Bradlyn,Wieder}. Notably, it is found that the triply-degenerate nodal point (TNP) could exist without nonsymmorphic symmetries~\cite{Heikkila,WengTNP,ZZhu,Chang}. It has been predicted to exist in a few inversion-asymmetric materials, which may exhibit interesting effects in spectroscopic and transport properties~\cite{WengTNP,ZZhu,Chang}.

Much effort has been devoted to the search of materials that can host such novel band-crossings. So far, the candidate materials with nodal rings or TNPs are still limited, and more importantly, their linear energy range, in which the quasiparticles can exhibit the unique properties of the band-crossing, is typically quite small (much less than 1 eV), putting stringent conditions on experimental studies. Thus, there is urgent need to search for realistic topological metal materials with a large linear energy range.

In the present work, based on first-principles calculations and symmetry analysis, we report that the family of metal diborides MB$_2$ (M=Sc, Ti, V, Zr, Hf, Nb, Ta) are a new type of topological metals with a pair of nodal rings coexisting with a pair of TNPs near the Fermi level. Importantly, the identified nodal ring and TNP here are both different from those in previous examples. The nodal ring here is formed from four entangled bands, not just two bands in previous examples, hence we term it as a four-band nodal ring (FNR). As a remarkable consequence, FNR features Dirac-cone like surface states, in sharp contrast to the usual drumhead surface states associated with two-band nodal rings. Owing to the presence of inversion symmetry, the TNPs here also show features distinct from previous ones. Especially, when spin-orbit coupling (SOC) is included, the TNP here transforms into a Dirac point that is close to the borderline between the type-I and type-II Dirac point categories. We clarify the symmetry protection of these novel band-crossing features and construct effective models for describing the low-energy excitations. A notable advantage of these materials is the large linear energy range ($>$ 2 eV), which should facilitate the further experimental study. Furthermore, it has been demonstrated that this family of materials can exhibit a wide range of fascinating properties, such as superconductivity (MgB$_2$, TaB$_2$, and ZrB$_2$)~\cite{41,42,43,44}, super-hardness (OsB$_2$, IrB$_2$, and ReB$_2$)~\cite{45,46,47}, and also excellent thermoelectricity (MgB$_2$, and AlB$_2$)~\cite{48,49}. Thus our work reveals a promising platform for studying the interplay between these exotic properties and the topological band-crossings.

\begin{figure}[t]
\includegraphics[width=9cm]{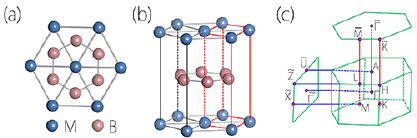}
\protect\caption{(a) Top and (b) side view of the crystal structure of MB$_2$ materials. (c) Brillouin zone with high symmetry points labeled.
} \label{fig1}
\end{figure}

\section{Crystal structure and methods}

The family of metal diborides MB$_2$ (M=Sc, Ti, V, Zr, Hf, Nb, Ta) share a layered hexagonal AlB$_2$-type structure with the space group of P6/mmm (No.~191), as illustrated in Fig.~\ref{fig1}(a,b). The boron atoms are arranged in layers, with layers of metal atoms M interleaved between them. The boron layer has a honeycomb structure, while the M layer has a hexagonal structure. The atomic positions are 1 M at (0,0,0) and 2 B at (1/3,2/3,1/2), (2/3,1/3,1/2). The crystal structure possesses a point group symmetry of $D_{6h}$. To be noted, all the MB$_2$ compounds studied in this work are existing materials that have been synthesized in experiment.

First-principles calculations based on the density functional theory (DFT), as implemented in the Vienna ab initio simulation package (VASP)~\cite{50} were performed to investigate the properties of these materials. The ionic potentials were treated by the projector augmented wave (PAW) approach~\cite{PAW}. We adopted the generalized gradient approximation (GGA) with the realization of Perdew-Burke-Ernzerhof (PBE) functional for the exchange-correlation potential~\cite{51}. The cutoff energy is chosen as 500 eV and the Brillouin zone was sampled with a 21$\times$21$\times$21  $\Gamma$-centered $k$-mesh for geometry optimization and for self-consistent calculations. The energy convergence criterion is set to be 10$ ^{-5}$ eV. The crystal structure is fully relaxed until the maximum force on each atom was less than 0.001 eV/\AA. The surface states are investigated by constructing the maximally localized Wannier functions~\cite{53,54} using the Wannier$_-$tools package~\cite{52} combined with an iterative Green's function method~\cite{55}.

\begin{figure}[t]
\includegraphics[width=9cm]{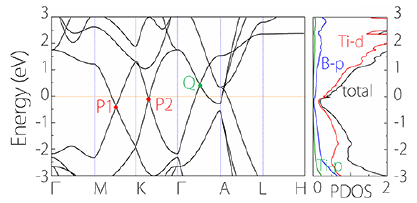}
\protect\caption{Calculated electronic band structure of TiB$_2$ without SOC and the projected density of states (PDOS). The three crossing points on the M-K, K-$\Gamma$, and $\Gamma$-A paths are labeled as $P_1$, $P_2$ and $Q$, respectively.
} \label{fig2}
\end{figure}

\section{Electronic band structure}

The electronic band structure of TiB$_2$ without SOC is displayed in Fig.~\ref{fig2} along with the projected density of states (PDOS). One observes that there are the highly dispersive bands at low energy, and several linear band-crossing points near the Fermi level can be identified. From the PDOS, it can be seen that the total DOS is small around Fermi level, showing the feature of a semimetal state. The low-energy states are dominated by the $d$-orbitals of the Ti atom. Of the band-crossing points in Fig.~\ref{fig2}, we will pay special attention to the following three: points $P_1$ and $P_2$ near the K point of the Brillouin zone, and point $Q$ on the $\Gamma$-A path. We shall see that $P_1$ and $P_2$ are actually points on a nodal ring, whereas point $Q$ is a TNP. Before proceeding, we comment that energy range of linear dispersion around these crossing points is almost more than 2 eV, much lager than other proposed materials with similar band features. This makes the current material a promising candidate for investigating the novel physics associated with these band-crossings in experiment.
\begin{figure}[t]
\includegraphics[width=8.5cm]{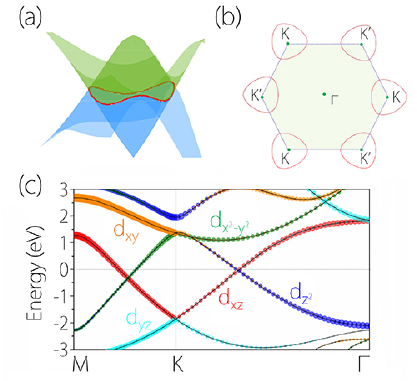}
\protect\caption{(a) Band dispersion around K point in the $k_z$=0 plane showing the nodal ring. (b) Illustration of the pair of nodal rings in the $k_z$=0 plane. (c) Band structure of TiB$_2$ near the K point. The size of the colored circles is proportional to the weight of projection onto atomic orbitals.
} \label{fig3}
\end{figure}

\subsection{Four-band nodal ring}

Let's first consider the crossing points $P_1$ and $P_2$. Noticing that the two points are in the $k_z=0$ plane, we perform a careful scan of the band structure around them. Interestingly, it shows that the two points are not independent, they are in fact residing on a nodal ring centered at K point in the $k_z=0$ plane. The energy dispersion around the ring is plotted in Fig.~\ref{fig3}(a), which clearly shows that the crossing is of linear-type and there is slight energy variation along the ring. As required by time reversal symmetry, there is another ring around the K' point, so totally there are a pair of inequivalent rings in the Brillouin zone (see Fig.~\ref{fig3}(b)).

There are two independent mechanisms that protect the nodal ring for the current system in the absence of SOC. First, the crystal has a horizontal mirror plane, and the rings are located in the mirror-invariant $k_z=0$ plane. Then each ring can be protected by the mirror symmetry if the two crossing-bands have different mirror eigenvalues, which is indeed the case as we have checked. Second, the system has inversion symmetry and time reversal symmetry, which requires that the Berry phase around a close loop in $k$-space must be quantized in unit of $\pi$~\cite{Ludwig}. For a small loop encircling each nodal ring, the Berry phase is given by $\pm \pi$, ensuring the protection of the ring. The two mechanisms are independent, meaning that by breaking one of them, the remaining one can still protect the ring from gap opening~\cite{Lu2016}.

\begin{figure}[t]
\includegraphics[width=7cm]{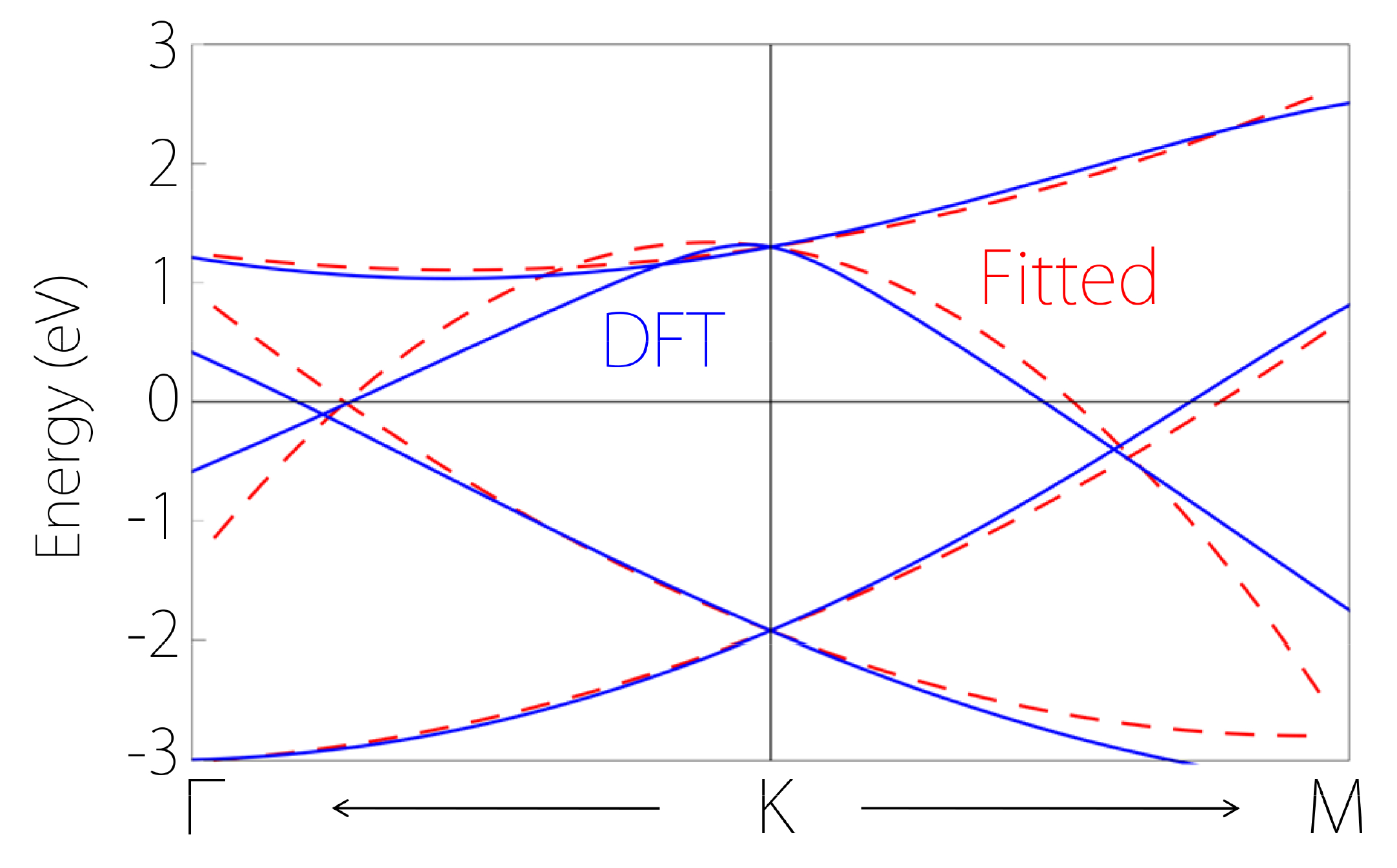}
\protect\caption{Fitted band structure (red dashed line) using the effective model compared with the DFT band structure (blue solid line).
} \label{fig4}
\end{figure}
Back to the band structure in Fig.~\ref{fig2}, one observes a distinct feature of the nodal ring here: its formation involves four entangled bands as more clearly seen in Fig.~\ref{fig3}(c), although the crossing at low energy is always between two bands. By projecting the Bloch states to atomic orbitals, one can see that the two crossing bands along M-K and K-$\Gamma$ are of different orbital characters, a clear indication of the strong mixing between the four bands. At K point, the four bands are degenerated into two pairs as dictated by symmetry (see Fig.~\ref{fig3}(c)), corresponding to the two-dimensional irreducible representations $E'$ and $E''$ of the $D_{3h}$ symmetry. In fact, using them as basis states, one can construct a $k\cdot p$ model to capture of the four-band nature of the ring. Up to the quadratic order, we obtain the $4\times 4$ effective Hamiltonian
\begin{equation}\label{FNRm}
\mathcal{H}=\left[
              \begin{array}{cc}
                h_{11} & h_{12} \\
                h_{12}^\dagger & h_{22} \\
              \end{array}
            \right],
\end{equation}
where each entry is a $2\times 2$ sub-matrix with
\begin{equation}
h_{ii}=M_i+D_ik_z^2+E_i k_\bot^2+\left(
              \begin{array}{cc}
                0 & A_i k_- +B_i k_+^2 \\
                A_i k_+ +B_i k_-^2 & 0 \\
              \end{array}
            \right),
\end{equation}
\begin{equation}
h_{12}=\left(
              \begin{array}{cc}
                A_3 k_z & B_3 k_z k_- \\
                B_3 k_z k_+ & A_3 k_z \\
              \end{array}
            \right).
\end{equation}
Here the wave vector $k$ is measured from K point, $i=1,2$, $k_\bot=\sqrt{k_x^2+k_y^2}$, $k_\pm= k_x\pm ik_y$, and the coefficients $M_i$, $D_i$, $E_i$, $A_i$, are $B_i$ are material-specific, and can be obtained by fitting the DFT band structure. Note that in the $k_z=0$ plane, $h_{12}$ vanishes, and $\mathcal{H}$ are decoupled into two $2\times 2$ diagonal blocks. Each block corresponds to a specific mirror eigenvalue with $+ 1$ ($-1$) for $h_{11}$ ($h_{22}$), consistent with our previous analysis that the two-bands crossing at the nodal ring have opposite mirror eigenvalues.  The fitted band structure along with the DFT result are plotted in Fig.~\ref{fig4}, which shows a good agreement.
A unique character that distinguishes FNR from conventional two-band nodal ring is that the former features Dirac-cone-type surface states, in contrast to the usual drumhead surface states of the latter. This can be directly derived from the effective model in Eq.~(\ref{FNRm}). Consider the (001) surface of the system, and we take the system to be semi-infinite in the region with $z>0$. Then the possible surface states can be obtained by solving the Schr\"{o}dinger equation with $\mathcal{H}$ and with the replacement $k_z\rightarrow -i\partial_z$. Straightforward calculation (see Appendix B) shows that there are two degenerate surface states at surface \={K} point (with degeneracy protected by the three-fold rotation $c_{3z}$ symmetry), and an effective Hamiltonian for the surface states can be obtained as
\begin{equation}
H_\text{surf}=v_F(k_x\sigma_x+k_y\sigma_y),
\end{equation}
which takes the standard form for a 2D massless Dirac fermion like that for graphene. Here, the Pauli matrices $\sigma_i$ stand for an orbital degree of freedom, and the Fermi velocity $v_F=(A_1+A_2)/2$. The Hamiltonian for \={K}' point will have an extra minus sign before $k_x$, as required by the time reversal symmetry. Thus, on the (001) surface, there must exist Dirac-cone-type surface states inside each surface-projected FNR.
\begin{figure}[t]
\includegraphics[width=8.5cm]{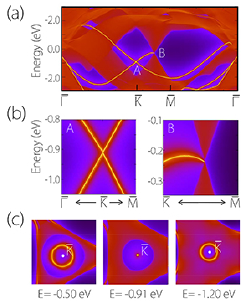}
\protect\caption{(a) Spectrum of TiB$_2$ (100)-surface. The surface Dirac cone at \={K} point can be clearly observed. (b,c) Enlarged view of two regions A and B marked in (a). (c) Calculated equi-energy slices near the \={K} point, exhibiting the Dirac-cone structure. The slice at $-0.91$ eV intersects the surface Dirac point at \={K} point.
} \label{fig5}
\end{figure}
This prediction is confirmed by our DFT result. Figure~\ref{fig5} shows the calculated spectrum of TiB$_2$ (001) surface. A surface Dirac-cone centered at \={K} point can be clearly observed, which is distinct from the usual drumhead surface states observed for two-band nodal rings. Topological insulators (TIs) also host Dirac-cone surface states. We point out that the Dirac-cone here is distinct from those for TIs in at least two aspects. (i) For TI, the surface Dirac cone is centered at time-reversal-invariant-momenta~\cite{Hasan2010,Qi2011}; whereas in our case, the cone is centered at \={K} and \={K}', which are not invariant under time reversal operation. (ii) For (strong) TI, each surface must have an odd number of Dirac cones~\cite{Hasan2010,Qi2011}; whereas in our case, there are two Dirac cones for (001) surface. Thus, the Dirac-cone surface states here is indeed a signature of FNR, distinct from other topological states. In addition, we note that a large part of the surface Dirac-cone is below the Fermi energy, making it readily detectable in ARPES experiment.
\begin{figure}[t]
\includegraphics[width=8.5cm]{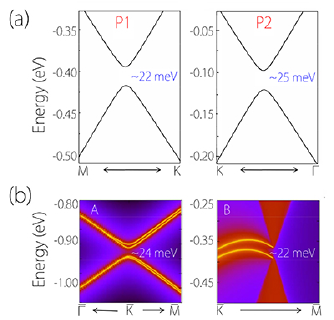}
\protect\caption{(a) Band structure around $P_1$ and $P_2$ with SOC included. The size of SOC-induced band gaps are indicated in the figures. (b) The spectrum for regions A and B as in Fig.~5(b) but with SOC included.
} \label{fig6}
\end{figure}
The above analysis is done without considering SOC. By including SOC, due to the presence of inversion and time reversal symmetries, each band will be doubly degenerate. The essential features of the band structure remain the same. The main effect of SOC is the opening of a small gap at the original band-crossing point (see Fig.~\ref{fig6}). For TiB$_2$, we find that the SOC gap size is about $22\sim25$ meV for the bulk FNR, while it is about 24 meV at the surface Dirac point (for (001) surface).

\subsection{Triply-degenerate nodal point}

\begin{figure}[t]
\includegraphics[width=8.5cm]{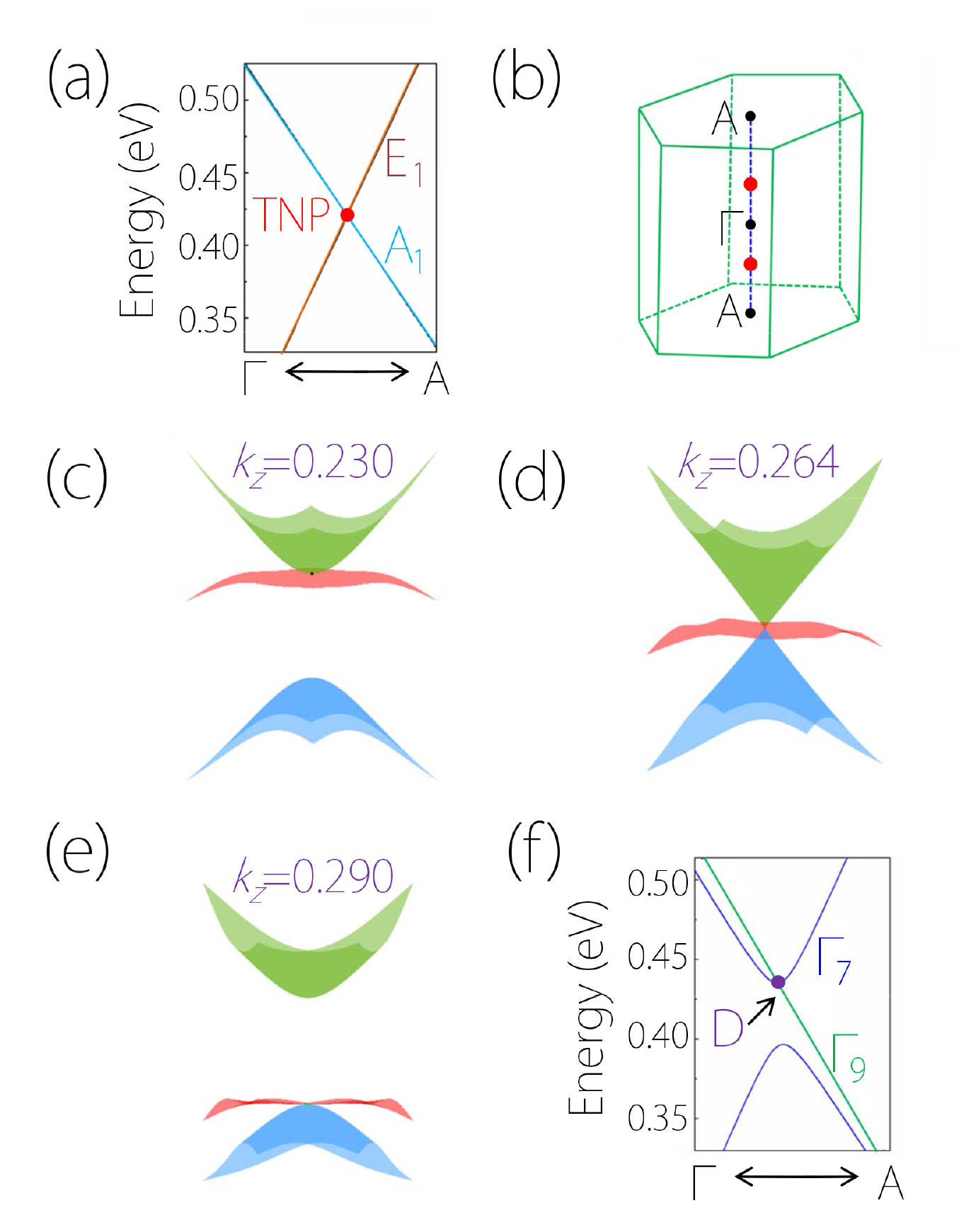}
\protect\caption{(a) The enlargement of electronic band structure around TNP of TiB$_2$. The irreducible representation of bands are indicated. (b) The position of TNP in the 3D bulk Brillouin zone. (c), (d) and (e) are the band dispersion in the $k_x$-$k_y$ plane for a constant $k_z$ value set above, at, and below the TNP, respectively. (f)Similar with (a), but for the electronic band structure along the $\Gamma$-A path with SOC included.
} \label{fig7}
\end{figure}

Now let's consider the $Q$ point along the $\Gamma$-A path as in Fig.~\ref{fig2} and Fig.~\ref{fig7}(a). Without SOC, one notes that the red-colored band is doubly-degenerate while the blue-colored band is nondegenerate (see Fig.~\ref{fig7}(a)). They respectively correspond to the $E_1$ and $A_1$ irreducible representations of the $C_{6v}$ symmetry group for the $\Gamma$-A line. Hence the crossing point between them is triply-degenerate. Due to time reversal symmetry, there is a pair of such TNPs, located on the two sides of the $\Gamma$ point, as illustrated in Fig.~\ref{fig7}(b).

In Fig.~\ref{fig7}(c-e), we plot the band dispersion in the $k_x$-$k_y$ plane for a constant $k_z$ value around the TNP. For a horizontal plane exactly intersecting the TNP (see Fig.~\ref{fig7}(d)), the top and bottom bands form a Dirac-cone-like bulk spectrum, while the middle band lies in-between and crosses the Dirac point, making the point triply-degenerate. For planes below (Fig.~\ref{fig7}(c)) or above (Fig.~\ref{fig7}(e)) the TNP, two of the three bands still stick together with a doubly-degenerate point along the $k_z$-axis, while the remaining one is detached. One notes that since the ordering between the $E_1$ band and the $A_1$ band is inverted between $\Gamma$ and A, they must cross each other in-between and form the TNP.

Using the $A_1$ and $E_1$ states at the TNP as basis, the low-energy effective model (up to linear order) around the TNP can be obtained as
\begin{equation}\label{TNPm}
\mathcal{H}=Ck_z+\left[
                   \begin{array}{ccc}
                     Ak_z & Bk_+ & Bk_- \\
                     Bk_- & 0 & 0 \\
                     Bk_+ & 0 & 0 \\
                   \end{array}
                 \right].
\end{equation}
Here the wave vector $k$ is measured from the TNP, and the energy of the TNP is set as zero energy. This model describes the low-energy quasiparticle excitations around the TNP.

TNPs were also reported in a few non-centrosymmetric materials~\cite{WengTNP,ZZhu,Chang}. The TNPs identified here share similar features as those examples. However, the presence of inversion symmetry in the current system makes a difference when SOC is considered. This is because without inversion symmetry, SOC will generally lift the band degeneracy, making each band nondegenerate; in contrast, the presence of inversion symmetry requires each band to be at least doubly degenerate. As shown in Fig.~\ref{fig7}(f), after including SOC, the original TNP is gapped. Interestingly, the two upper bands now cross at a single point $D$ with four-fold degeneracy. This point is symmetry-protected, because the two crossing bands are of different irreducible representations ($\Gamma_7$ and $\Gamma_9$) of the $C_{6v}$ double group. Again, using the symmetry character of the bands, we obtain the effective model expanded around the crossing-point:
\begin{equation}
\mathcal{H}_D=\left[
                \begin{array}{cc}
                  h_+ & 0 \\
                  0 & h_- \\
                \end{array}
              \right],
\end{equation}
where each entry is a $2\times 2$ block, with
\begin{equation}
h_\pm =Ck_z+A(k_x\sigma_x\pm k_y\sigma_y)+Bk_z\sigma_z.
\end{equation}
Note that the coefficients ($A$, $B$ and $C$) here are different from those in Eq.(\ref{TNPm}).
In the expression of $h_\pm$, the first term is an overall energy tilt, while the remain terms is just that for a Weyl point with $\pm$ chirality. Thus, the four-fold crossing point consists of two Weyl points with opposite chirality, making it a Dirac point. A notable feature of this Dirac point is that one of the crossing band ($\Gamma_7$) has almost zero slope at $D$ along $k_z$ direction (see Fig.~\ref{fig7}(f)), because $|C|\approx|B|$. A Dirac point can be classified as type-I (type-II) when the two crossing bands have opposite (the same) sign of slope along any (certain) direction~\cite{Type-II56,Type-II57}. The current case is near the borderline between type-I and type-II. Such band crossing features have not been observed in previous works on TNPs.

\section{Discussion and Conclusion}

We take TiB$_2$ as a representative in the above presentation. As shown in Appendix A, the essential features, including the FNRs and TNPs, are also shared by other members of this material family. In some materials, like ScB$_2$ and VB$_2$, the FNR appears above the Fermi energy, whereas in others, the FNR is below the Fermi energy. For the TNP, it is above Fermi energy for most members except for NbB$_2$ and TaB$_2$.

Experimentally, the novel band features predicted here can be probed by the ARPES technique. Particularly, the Dirac-cone surface state associated with the FNR should be readily detected. As for the TNPs, previous works have shown that under magnetic field, there appears anomalous chiral Landau levels~\cite{WengTNP,ZZhu,Chang}. This should manifest in magneto-transport experiment as unusual field-dependent magneto-resistance, when both $E$ and $B$ fields are along the $z$-direction ($c$-axis).

In conclusion, we have predicted novel types of topological metals in a family of metal diboride materials. These materials feature a pair of FNRs coexisting with a pair of TNPs near the Fermi level. The FNR is distinct from the previously discussed two-band nodal rings in that its formation necessarily requires four entangled bands, and it leads to Dirac-cone-like surface states rather than the usual drumhead surface states for two-band nodal rings. The Dirac-cone-like surface states here are also distinct from those for the three-dimensional topological insulators, as we have discussed. Such kind of FNRs is discovered here for the first time. The TNPs in the current system also exhibit different features from the previously reported ones. Due to the presence of inversion symmetry, under SOC, the TNP transforms into a Dirac point that is near the borderline between type-I and type-II Dirac point categories. The large linear energy range associated with these band crossings will facilitate the experimental detection of their interesting properties. Given that these materials exhibit a wide range of fascinating properties, such as superconductivity, superhardness, and excellent thermoelectricity, it will be highly interesting to explore the interplay between these nice properties and the topological band structure in the future.

\begin{acknowledgments}
The authors thank D.L. Deng for helpful discussions. This work is supported by Singapore Ministry of Education Academic Research Fund Tier 2 (MOE2015-T2-1-150) and Tier 1 (SUTD-T1-2015004). 
\end{acknowledgments}

\begin{figure*}[t]
\includegraphics[width=16cm]{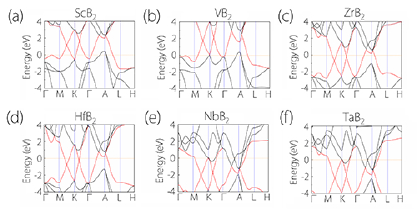}
\protect\caption{(a)-(f) Electronic band structure of MB$_2$ (M= Sc, V, Zr, Hf, Nb, Ta) materials without SOC.
} \label{fig8}
\end{figure*}

\begin{appendix}
\section{Band Structure OF MB$_2$ Family}

Fig.8 shows the electronic band structures of the MB$_2$ (M=Sc, V, Zr, Hf, Nb, Ta) family of materials (in the absence of SOC). One observes similar features as those discussed in the main text for TiB$_2$. Particularly, the band-crossings along M-K and K-$\Gamma$ (for the FNR) and the TNP along $\Gamma$-A can be observed.

\section{Surface Dirac cone for FNR}
Consider the system described by the low-energy model Eq.(\ref{FNRm}) occupying the half space with $z>0$.
In such case, while $k_x$ and $k_y$ are still good quantum numbers, $k_{z}$ is not, hence we need to replace $k_z$
by $-i\partial_{z}$ in the low-energy Hamiltonian. Since we shall focus on the dispersion near the surface \={K} point (where $k_x$ and $k_y$ are small), we separate the Hamiltonian into two parts: $\mathcal{H}=\mathcal{H}_{0}+\mathcal{H}_{1}$,
with $\mathcal{H}_{0}=\mathcal{H}(k_{x}=0,k_{y}=0)$, and $\mathcal{H}_{1}$ is the remaining part of which the terms at least linear in the small parameters $k_x$ and $k_y$.

Via a unitary transformation, $\mathcal{H}_{0}$ can be put into a block-diagonal form, such that the Schr\"{o}dinger equation for states at \={K} point reads
\begin{eqnarray}
\tilde{\mathcal{H}}_{0}\Psi(z) & = & \left[\begin{array}{cc}
\tilde{H}_{z}(-i\partial_{z}) & 0\\
0 & \tilde{H}_{z}(-i\partial_{z})
\end{array}\right]\Psi(z)=E\Psi(z), \nonumber \\
\end{eqnarray}
with
\begin{eqnarray}
\tilde{H}_{z}(-i\partial_{z}) & = & M_{+}+\left(\begin{array}{cc}
M_{-} & -iA_3 \partial_{z}\\
-iA_3 \partial_{z} & -M_{-}
\end{array}\right),
\end{eqnarray}
with $M_{\pm}=\frac{1}{2}[M_1\pm M_2-(D_{1}\pm D_{2})\partial_{z}^{2}]$.
Since the two blocks are identical, the eigenstates at surface \={K} point must be at least doubly degenerate. The eigenstates are of the form
\begin{eqnarray*}
\Psi_{1}(z)=\left(\begin{array}{c}
\psi_{0}\\
\boldsymbol{0}
\end{array}\right), & \ \  & \Psi_{2}(z)=\left(\begin{array}{c}
\boldsymbol{0}\\
\psi_{0}
\end{array}\right),
\end{eqnarray*}
where $\boldsymbol{0}=(0,\ 0)^{T}$. To solve for possible boundary solutions, we take the ansatz
\begin{eqnarray}
\psi_{0} & = & \left(ae^{\lambda_{+}z}+be^{\lambda_{-}z}\right)\phi_{+}+\left(ce^{-\lambda_{+}z}+de^{-\lambda_{-}z}\right)\phi_{-},\nonumber \\
\end{eqnarray}
where the spinor part $\phi_\pm$ satisfies $\sigma_{y}\phi_{\pm}=\pm\phi_{\pm}$ and $\lambda_{\pm}=\frac{1}{D_{2}-D_{1}}[A_3\pm\sqrt{A_3^{2}+(M_{1}-M_{2})(D_{1}-D_{2})}]$.
The boundary condition $\psi_{0}(0)=0$, together with the finiteness
of wave-function, lead to
\begin{eqnarray}
\psi_{0}(z) & \propto & \begin{cases}
\left(e^{\lambda_{+}z}-e^{\lambda_{-}z}\right)\phi_{+}, & \frac{A_3}{D_{2}-D_{1}}<0,\\
\left(e^{-\lambda_{+}z}-e^{-\lambda_{-}z}\right)\phi_{-}, & \frac{A_3}{D_{2}-D_{1}}>0.
\end{cases}
\end{eqnarray}
For ${\rm TiB_{2}}$, we have  $A_3/(D_{2}-D_{1})>0$ from the fitting result, and hence
the two doubly-degenerate states at the \={K} point are
\begin{eqnarray}
&&\Psi_{1}(z)=\frac{1}{N}\left(e^{-\lambda_{+}z}-e^{-\lambda_{-}z}\right)\left(\begin{array}{c}
\phi_{-}\\
\boldsymbol{0}
\end{array}\right),  \\
&& \Psi_{2}(z)=\frac{1}{N}\left(e^{-\lambda_{+}z}-e^{-\lambda_{-}z}\right)\left(\begin{array}{c}
\boldsymbol{0}\\
\phi_{-}
\end{array}\right),
\end{eqnarray}
where $N$ is the normalization constant.
Then the effective Hamiltonian for surface states around \={K} point are obtained as~\cite{Shen}
\begin{equation}
H_\text{surf}=\left(\begin{array}{cc}
\langle\Psi_{1}|\tilde{\mathcal{H}}_{1}|\Psi_{1}\rangle & \langle\Psi_{1}|\tilde{\mathcal{H}}_{1}|\Psi_{2}\rangle\\
\langle\Psi_{2}|\tilde{\mathcal{H}}_{1}|\Psi_{1}\rangle & \langle\Psi_{2}|\tilde{\mathcal{H}}_{1}|\Psi_{2}\rangle
\end{array}\right),
\end{equation}
To the linear order in $k_{x}$ and $k_{y}$, we obtain
\begin{equation}
H_\text{surf}=\frac{A_1+A_2}{2}(k_{x}\sigma_{x}+k_{y}\sigma_{y}),
\end{equation}
which is a surface Dirac-cone inside the projected FNR.
\end{appendix}

\end{document}